\documentclass[aip,reprint,amsmath,amssymb]{revtex4-2}

\makeatletter
\def\section{\@startsection{section}{1}{\z@}{-2.2ex \@plus -0.5ex \@minus -.2ex}{1.0ex \@plus .2ex}{\normalfont\normalsize\bfseries}}
\def\subsection{\@startsection{subsection}{2}{\z@}{-1.8ex \@plus -0.5ex \@minus -.2ex}{0.7ex \@plus .2ex}{\normalfont\normalsize\bfseries}}
\def\subsubsection{\@startsection{subsubsection}{3}{\z@}{-1.5ex \@plus -0.5ex \@minus -.2ex}{0.5ex \@plus .2ex}{\normalfont\normalsize\bfseries}}
\makeatother

\usepackage{amsmath,amssymb}
\usepackage{braket}
\usepackage{booktabs}
\usepackage{graphicx}
\usepackage{hyperref}
\usepackage{xcolor}
\usepackage{mhchem}
\usepackage{braket}
\usepackage{inconsolata}
\usepackage[T1]{fontenc}
\usepackage{enumitem}
\setlist[itemize]{leftmargin=*, labelsep=0.4em, topsep=3pt, itemsep=3pt, parsep=0pt}
\newcommand{\code}[1]{\texttt{#1}}

\def\ket#1{| #1 \rangle}
\def\bra#1{\langle #1 |}

\def\int#1#2{\left\langle #1 | #2 \right\rangle}
\def\aint#1#2{\bra{#1}\ket{#2}}

\begin{document}

\title{PyCC: A Reference Quantum Chemistry Python Package for Education, Validation, and Rapid Prototyping}
\author{T. Daniel Crawford}
	\affiliation{Department of Chemistry, Virginia Tech, Blacksburg, VA 24061, U.S.A.}
	\email{crawdad@vt.edu}
\author{Benjamin G. Peyton}
    \altaffiliation[Present address: ]{Department of Chemistry, Michigan State University, East Lansing, MI 48824, U.S.A.}
	\affiliation{Department of Chemistry, Virginia Tech, Blacksburg, VA 24061, U.S.A.}
\author{Zhe Wang}
    \altaffiliation[Present address: ]{School of Chemistry and Biochemistry, Georgia Institute of Technology, Atlanta, GA
        30332-0400, U.S.A.}
	\affiliation{Department of Chemistry, Virginia Tech, Blacksburg, VA 24061, U.S.A.}
\author{Aparna Krishnan}
	\affiliation{Department of Chemistry, Virginia Tech, Blacksburg, VA 24061, U.S.A.}
\author{Brendan M. Shumberger}
	\affiliation{Department of Chemistry, Virginia Tech, Blacksburg, VA 24061, U.S.A.}
\author{Jose Madriaga}
    \altaffiliation[Present address: ]{Department of Physics, University of Texas at Dallas, Richardson, TX 75800, U.S.A.}
	\affiliation{Department of Chemistry, Virginia Tech, Blacksburg, VA 24061, U.S.A.}
\author{Joshua Atta-Kumi}
    \altaffiliation[Present address: ]{Nesvard Institute of Molecular Sciences, Accra, Ghana}
	\affiliation{Department of Chemistry, Virginia Tech, Blacksburg, VA 24061, U.S.A.}
\author{Aaron De Clercq}
    \altaffiliation[Present address: ]{Helmholtz Institute Erlangen-N\"urnberg, 91058 Erlangen, Germany}
	\affiliation{Department of Chemistry, Virginia Tech, Blacksburg, VA 24061, U.S.A.}
\author{Ritvik Prabhu}
    \altaffiliation[Present address: ]{Department of Computer Science, Virginia Tech, Blacksburg, VA 24061, U.S.A.}
	\affiliation{Department of Chemistry, Virginia Tech, Blacksburg, VA 24061, U.S.A.}
\date{\today}

\begin{abstract} PyCC is a Python-based quantum chemistry package that
emphasizes coupled cluster and related methods.  Developed over the last five
years, PyCC's purpose is to provide a clear reference implementation that will
serve as a validation suite for production code, a platform for rapid
prototyping and testing of new methods, and an educational resource for
newcomers to the field. The package covers both closed- and open-shell
single-determinant references in spatial- and spin-orbital formulations for
the coupled cluster hierarchy through triples-including methods such as CC3,
as well as for MP2 and CI methods. PyCC provides a broad range of molecular
properties such as analytic first and second energy derivatives including
gradients, Hessians, polarizabilities and atomic polar tensors, as well as
non-energy-derivative properties, such as velocity-gauge atomic polar tensors
and diagonal Born-Oppenheimer corrections, plus dynamic linear response
functions, and excitation energies.  The package includes a number of unique
capabilities, such as real-time CC propagation (up through CC3), atomic axial
tensors for simulating vibrational circular dichroism spectra at the MP2,
CID, CISD, and CCSD levels, and local correlation methods including perturbation-aware
correlation domains. The code is designed around a shared \code{Wavefunction}
base and a method-agnostic derivative layer that enables a new method to
inherit the full energy-derivative property set once it supplies its reduced
densities and their first-order responses.  PyCC is fully open source and
available on GitHub under the BSD 3-clause license.  \end{abstract}

\maketitle

\section{Introduction}

The inherent complexity of wave-function-based quantum chemical methods such
as coupled cluster (CC) theory\cite{Bartlett05, Bartlett07, Bartlett10,
Bartlett2024, Crawford2026:handbook, Shavitt09, Crawford00:review} is a
challenge to the development of highly optimized production-level software
implementations.  Even at moderate levels of electron correlation (up to
triple excitations), the CC amplitude equations alone involve dozens of tensor
contractions,\cite{Kallay01} and careful validation and testing at each stage
of development is essential for robust, trustworthy code.  In addition, while
optimized code is naturally highly sophisticated, it is typically accessible
only to experts and thus presents a large learning curve for students and
other newcomers to the field. Simpler, well-documented reference codes can be
invaluable for exploratory and/or educational purposes.

Over the last five years we have developed PyCC, a Python-based coupled
cluster implementation intended to serve as an educational resource, a
platform for rapid prototyping of new methods, and a validation suite for
production-level codes.  As the name implies, the package focuses primarily on
CC theory, but we have implemented other methods (e.g., M{\o}ller-Plesset
perturbation theory and configuration interaction) for specific target
applications.\cite{Shumberger2024, Shumberger2025, Shumberger2026, Krishnan2026:density}  By design, the code
emphasizes closed-shell, spin-adapted references and provides a range of CC
capabilities, including ground- and excited-state energies, analytic first and
second derivatives, linear and quadratic response functions (with the latter
under pull-request review), and explicitly time-dependent wave functions. In
addition, portions of the code are GPU-capable, and users can select
between single and double arithmetic precision for both energies and
time-dependent methods.  Furthermore, the code provides a convenient testbed
for reduced-scaling models, such as local-pair-natural-orbital (LPNO) and
related approaches.

In this paper we discuss the details of the PyCC code, starting in the next
section with the overall structure and design philosophy followed by
highlights of specific implementation aspects for representative methods,
including GPU and reduced-scaling methods.  Finally, we discuss details of
certain unique capabilities of PyCC, such as time-dependent CC including
perturbative triples, LPNO-based response functions, and the atomic axial
tensors required for vibrational circular dichroism.

\section{Current Capabilities}

PyCC implements a broad set of ground-state, response and time-dependent electronic-structure
methods, all built on a single \texttt{Wavefunction} base class.  This section surveys what the
code can do at present, whereas the formulation and implementation of individual capabilities are taken up
in the sections that follow.

\begin{itemize}

\item \textbf{References and correlation methods.}  Restricted, unrestricted,
and restricted open-shell Hartree-Fock references (RHF, UHF,
ROHF)\cite{Szabo89} are supported, in either a spin-adapted closed-shell
formalism or a spin-orbital formalism, with the latter providing the
open-shell correlated methods.  The available electron-correlation methods
are: second-order M{\o}ller--Plesset perturbation theory
(MP2),\cite{MollerPlesset34} configuration interaction\cite{Sherrill98:CI}
with doubles and with singles and doubles (CID, CISD), and the coupled-cluster
hierarchy CCD,\cite{Bartlett78, Pople78} CC2,\cite{Christiansen95:CC2}
CCSD,\cite{Purvis82} CCSD(T),\cite{Raghavachari89} and CC3.\cite{Koch97}
Ground-state energies are available for all three references, apart from CCD
and CC2, which are restricted to closed-shell RHF.  Frozen-core orbitals are
supported throughout except in the reduced-scaling methods.

\item \textbf{Ground-state machinery.}  For CC methods, the
similarity-transformed Hamiltonian ($\bar{H} = e^{-\hat{T}} \hat{H}
e^{\hat{T}}$) and (left-hand) $\Lambda$ amplitudes are available for most
methods. In addition, one- and two-particle reduced density matrices are
available using RHF, UHF and ROHF references for MP2, CCSD and CCSD(T), and
for RHF references for CID and CISD.  (One-electron densities are also
available for CC3.) These densities feed into the property, response, and
real-time CC codes.

\item \textbf{Analytic derivatives and static properties.}  Nuclear energy
gradients, molecular Hessians (force-constant matrices), electric dipole
moments (relaxed and unrelaxed), static dipole polarizabilities, length- and velocity-gauge atomic polar
tensors (LG-APTs and VG-APTs), atomic axial tensors (AATs), and diagonal Born--Oppenheimer corrections (DBOCs),
summarized in Table~\ref{tab:derivatives}.  All energy-derivative methods are built on shared
orbital-response (coupled-perturbed Hartree-Fock and Z-vector) and $2n+1$
machinery,\cite{Yamaguchi94, Handy84} such that a new method inherits the full
set once its reduced densities and their first-order responses are supplied.
A harmonic vibrational analysis function accepts these quantities directly to
produce frequencies, normal modes, and infrared intensities and vibrational
circular dichroism rotatory strengths.

\begin{table*}
\caption{\label{tab:derivatives}Analytic derivative properties available in PyCC.  A bullet marks
an implemented property; a blank marks one that is not currently implemented.  The listed
self-consistent-field references apply to every property in the row.  All properties are available
with and without frozen-core orbitals, except in the Hartree--Fock row, which is all-electron by
construction.  The Hartree--Fock dipole moment is an expectation value, while the correlated dipole
moments and all of the static polarizabilities are orbital-relaxed and so require the
coupled-perturbed Hartree--Fock (Z-vector) orbital response.  APT and AAT denote the atomic polar
and atomic axial tensors, the intensity parameters of infrared and vibrational circular dichroism
spectra respectively, and DBOC denotes the diagonal Born-Oppenheimer energy correction.}
\begin{ruledtabular}
\begin{tabular}{llcccccccc}
Method & References & Gradient & Hessian & Dipole & \shortstack{Static\\polarizability}
       & \shortstack{APT\\(length gauge)} & \shortstack{APT\\(velocity gauge)} & AAT & DBOC \\
\colrule
HF      & RHF, UHF, ROHF\footnotemark[1] & $\bullet$ & $\bullet$ & $\bullet$ & $\bullet$ & $\bullet$ & $\bullet$ &
$\bullet$ & $\bullet$ \\
MP2     & RHF, UHF & $\bullet$ & $\bullet$ & $\bullet$ & $\bullet$ & $\bullet$ & $\bullet$ & $\bullet$ & $\bullet$ \\
CISD    & RHF      & $\bullet$ & $\bullet$ & $\bullet$ & $\bullet$ & $\bullet$ & $\bullet$ & $\bullet$ & $\bullet$ \\
CCSD    & RHF, UHF & $\bullet$ & $\bullet$ & $\bullet$ & $\bullet$ & $\bullet$ & $\bullet$ & $\bullet$ & $\bullet$ \\
CCSD(T) & RHF, UHF & $\bullet$ & $\bullet$ & $\bullet$ & $\bullet$ & $\bullet$ &   N/A     &  N/A      & N/A \\
\end{tabular}
\footnotetext[1]{ROHF references are available for the gradient and the dipole moment, neither of
which requires an orbital response.  The remaining Hartree--Fock properties, and every correlated
property in this table, require the coupled-perturbed Hartree--Fock equations, which are not
implemented for ROHF references.}
\end{ruledtabular}
\end{table*}
 
\item \textbf{Frequency-dependent response.}  Coupled-cluster linear response
functions\cite{Helgaker12} are available to compute dynamic dipole polarizabilities and optical
rotation, for CCSD with RHF, UHF and ROHF references and for CC3 in the
spin-orbital formalism.  These are computed using the conventional
orbital-unrelaxed formulation and are therefore distinct from the
orbital-relaxed static polarizabilities of Table~\ref{tab:derivatives}.
However, these properties can be computed at the CCSD level through two
routes: a $2n+1/2n+2$ formulation for which only field-derivatives of the
ground-state $\hat{T}$ amplitudes are required at $+/-$ field frequencies
(referred to in PyCC as the symmetric approach), and a $2n+1$ formulation that
requires field-derivatives of both $\hat{T}$ and $\hat{\Lambda}$ amplitudes
(the asymmetric approach).

\item \textbf{Excited states.}  Equation-of-motion CCSD (EOM-CCSD) excitation
energies,\cite{Stanton93:EOM} for RHF, UHF and ROHF references.  PyCC can
compute multiple right- and/or left-hand excited-state eigenvectors starting
from several user-selected guess wavefunctions, including unit vectors, CIS
wave functions, or the eigenvectors of the singles-singles block of $\bar{H}$.

\item \textbf{Explicitly time-dependent coupled cluster.}  Real-time
propagation\cite{Ofstad2023} of the CCD, CC2, CCSD, and CC3 wave functions
under an external electric field, with a selection of numerical
integrators,\cite{Hairer93:ODE1, Hairer96:ODE2, Butcher16:ODE} several laser
pulse envelopes, and Pad\'e-accelerated spectra.  Available integrators
include seven fixed-step explicit Runge-Kutta integrators, one adaptive
embedded integrator (Cash-Karp),\cite{Cash90} and one implicit Gauss-Legendre
collocation integrator, with further embedded and implicit methods in
progress.

\end{itemize}

\subsection{Distinctive capabilities}

Three of the above are, to our knowledge, unique to PyCC and are treated in detail later in this paper.

\begin{itemize}

\item \textbf{Real-time coupled cluster, including RT-CC3.}  Explicit time propagation of the
coupled-cluster wave function at the CC3 level,\cite{Wang2025} in addition to CCD, CC2 and CCSD.

\item \textbf{Vibrational circular dichroism and related tensors.} Tensors
associated with certain properties require overlaps of wave-function
derivatives, such as the AATs\cite{Stephens85} and VG-APTs\cite{Amos88,Shumberger2026:LGOI} of VCD spectra and the 
DBOC. While VCD tensors have been available at
HF, DFT, and MCSCF levels of theory for decades,\cite{Amos87, Bak93,
Cheeseman96} PyCC includes our recently developed MP2, CID, CISD, and CCSD
methods,\cite{Shumberger2024, Shumberger2025, Shumberger2026, Crawford2026:ccvcd} including a
density-based formulation.\cite{Krishnan2026:density} 
PyCC also includes DBOCs, which use a similar
formulation and have been previously reported in the literature for MP2, CISD
and CCSD.\cite{Gauss2006, Tajti2007}

\item \textbf{Perturbation-adapted local correlation.}  Local correlation
spaces constructed from projected atomic orbitals (PAOs)\cite{Saebo93, Hampel96, Schutz01} and from pair natural
orbitals (PNOs),\cite{Neese09} including the perturbation-adapted PNO++ and CPNO++ spaces
built from perturbed pair densities,\cite{Crawford19, Dcunha2021, Madriaga2025} applied to CCSD and CCD
energies and to real-time propagation.\cite{Peyton2023}

\end{itemize}

\subsection{Comparison to Other Python-Based Codes}

A number of other Python packages that emphasize CC theory are available with
some overlap in purpose and capabilities with PyCC, most notably
PySCF,\cite{Sun2018:PySCF, Sun2020:PySCF} Psi4NumPy,\cite{Smith18}
CCpy,\cite{CCpy} Penguin,\cite{Hillers-Bendtsen2025} and HyQD.\cite{HyQD:CC}
The PySCF code provides near-production-level calculations with Python
interfaces, and includes CC-level energies and analytic energy first
derivatives, as well as EOM-CC, but is principally user-facing rather than a
reference implementation.  Psi4NumPy has a similar educational intent to PyCC,
but comprises scripts and notebooks rather than a full package.  (We note that
the \code{helper\_diis} code in PyCC was originally based on the corresponding
code in Psi4NumPy, with extensions to allow deployment on GPU systems.) The
CCpy package from the Piecuch group also focuses primarily on CC methods and
is much more ambitious than PyCC in its deployment of higher excitations.
Penguin has a similar purpose to CCpy, but also focuses on the development of
production-level implementations not only for CC, but also cluster
perturbation and M{\o}ller-Plesset perturbation theory hierarchies.  Finally,
the HyQD code from the Pedersen group overlaps with PyCC in its focus on RT-CC
methodology, and both PyCC and HyQD have benefited from collaborations between
our groups.  None of these packages includes CC analytic second derivatives or
the other features noted in the previous section.  Thus, PyCC is distinct from
all of these in part because of the breadth of CC-level properties it offers,
as well as its focus on both education and serving as a reference/validation
suite for on-going production-level codes, such as Psi4.

 \section{Design Philosophy and Code Structure}

The fundamental goals of PyCC are (1) to provide a clean, well-documented
reference implementation of a range of quantum chemical electron correlation
methods, with an emphasis on CC theory; (2) to serve as a validation suite for
the development of advanced production-level codes; (3) to allow rapid
prototyping of new ideas and methods; and (4) to facilitate the education of
students and other novices to the fundamental algorithmic infrastructure of
advanced quantum chemical methods.  With these goals motivating its
construction, PyCC fills an important role in research in quantum chemistry,
namely that of intermediate-level software that may be used for rapid
prototyping of new methods and models.  Particularly when implemented in
relatively simple yet multi-paradigm and full-featured languages such as
Python, such codes allow for preliminary, yet publication-quality
studies.\cite{Peyton2023, Madriaga2025, Madriaga2025:disc, Wang2025,
Krishnan2026} Furthermore, given the availability of robust
scientific-computing packages such as NumPy,\cite{Harris2020:numpy}
\code{opt\_einsum},\cite{Smith2018:opteinsum} SciPy,\cite{Virtanen2020:scipy}
PyTorch,\cite{Paszke2019:pytorch} \code{h5py},\cite{Collette2013:h5py, HDF5}
and others, such codes are not limited merely to ``toy'' problems and can
inform the development of highly optimized algorithms.

\subsection{The Wavefunction Base Class}
\label{subsec:wfn}

\begin{figure*}[!hbtp]
\centering
\includegraphics[width=\textwidth]{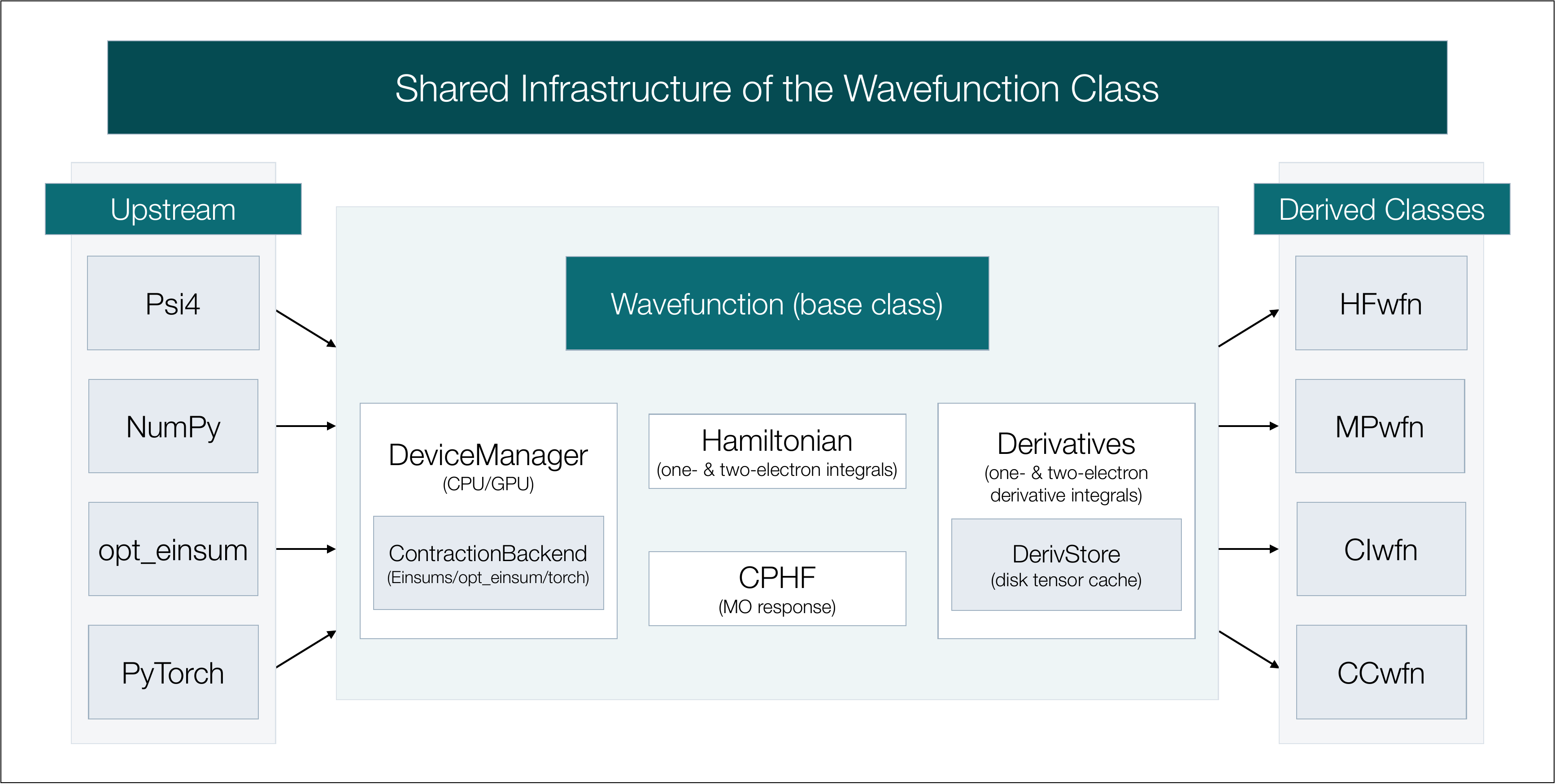}

\caption{\code{Wavefunction} base class overview. In addition to common
components described in the text, such as the reference wave function, the
\code{Wavefunction} object creates and uses several helper classes including a
\code{DeviceManager} for selecting CPU/GPU, arithmetic precision and
tensor-contraction handling (\code{ContractionBackend}), a
\code{Hamiltonian}/\code{SpinOrbitalHamiltonian} object for molecular
integrals and other quantities, \code{Derivatives} and \code{DerivStore}
objects for lazy construction and storage of derivative integrals and wave function
parameters, and a \code{CPHF} object for handling orbital response to external
perturbations.  Upstream packages such as Psi4 and NumPy provide essential
data or tools, while derived classes utilize the base class's attributes for
method-specific calculations.}

\label{fig:wfn_base_class} \end{figure*}

The \code{Wavefunction} class holds a set of common components used by every
quantum chemical method in PyCC.  The SCF reference wave function, \code{ref},
is computed by the Psi4 package\cite{Smith20} and provided to the constructor
by the user.  The \code{ref} provides \code{Wavefunction} with the molecular
orbital (MO) coefficients, the SCF energy, nuclear repulsion energy,
frozen-core data, orbital symmetries, and other attributes. While PyCC does
not make use of point-group symmetry in its calculations, it accepts MOs that
are either symmetrized or unsymmetrized.  In addition, the MO basis is
automatically resolved from the reference by default either as
\code{'spatial'} for closed-shell RHF references or \code{'spinorbital'} for
UHF/ROHF references, and these choices also yield key information about the number
of occupied/virtual orbitals and NumPy slices, spin-to-spatial-MO mappings,
etc.  Furthermore, the \code{Wavefunction} builds and maintains several key
helper classes, as depicted in Fig.~\ref{fig:wfn_base_class}.

\subsubsection{DeviceManager}

The \code{DeviceManager} controls the PyCC policies for the types of computing
device, the desired arithmetic precision, and the choice of tensor-contraction
engine.  For CUDA-based GPU systems, two device handles are automatically
assigned as \code{device0} for the CPU and \code{device1} for the GPU.  Such
calculations (currently limited to CC-level energy calculations and RT-CC
trajectories) require the PyTorch package but automatically fall back to
CPU-only when it is unavailable.  The arithmetic precision of the entire suite
of PyCC computations may be user-selected with \code{'SP'} for single
precision (float32/complex64) vs.\ \code{'DP'} for double precision
(float64/complex128), where the latter is the default.  (It should be noted,
however, that formal testing and validation of CPU/GPU capabilities and the
impact of SP/DP on numerical results have been carried out only for
ground-state energies and RT-CC methods.  Application to derivative properties
has not yet been investigated.) Finally, the \code{ContractionBackend} object
routes all tensor contractions through \code{opt\_einsum.contract} for both
CPU and GPU devices.  For the latter, the \code{ContractionBackend} first
moves any operand not already seeded onto \code{device1} and then upcasts real
operands to complex when the operand list is mixed.  This structure allows for
the top-level tensor-contraction calls throughout the PyCC code to avoid any
explicit references to devices and thus remain CPU/GPU agnostic.  The
structure also permits the seamless addition of improved tensor-contraction
functionality in the future, such as the Einsums C++ library currently under
development.\cite{Einsums}

\subsubsection{Hamiltonian and SpinOrbitalHamiltonian}

The \code{Hamiltonian} and \code{SpinOrbitalHamiltonian} objects use the Psi4
\code{MintsHelper} object\cite{Turney12:PSI4} to compute one- and two-electron
integrals and transform them into the spatial or spin-orbital basis,
respectively.  Both objects use certain common attribute names such as for the
Fock matrix (\code{F}), the MO energies (\code{eps}), the two-electron
repulsion integrals (\code{ERI}), various one-electron property integrals such
as electric-dipole moment (\code{mu}), magnetic-dipole moment (\code{m}),
linear momentum (\code{p}), and traceless quadrupole (\code{Q}).  The
\code{ERI} are stored in Dirac's notation as $\langle pq | rs \rangle$ for
spatial orbitals and in antisymmetrized form $\langle pq || rs \rangle$ for
spin orbitals.  The spin-orbital \code{ERI} tensor is populated using
spin-masked assignments from the corresponding spatial MO integrals, with no
Python loops over orbital quadruples.  In addition, the spatial-orbital
\code{Hamiltonian} contains a redundant set of two-electron integrals
(\code{L}) as $2 \langle pq | rs \rangle - \langle pq | sr \rangle$ to
streamline contraction expressions for spin-adapted wave functions.

For spin-orbital cases, while semicanonicalization of the orbitals is carried
out automatically, no assumption is made in any of the wave function
contraction codes that the orbitals are canonical (apart from CCSD(T) and CC3,
which require it to avoid storage of full triples amplitudes) or even that
Brillouin's theorem is obeyed (\textit{i.e.}, that the occupied-virtual block
of the Fock matrix is zero).  Furthermore, frozen core capabilities are
handled implicitly in all methods by storing the
\code{Hamiltonian}/\code{SpinOrbitalHamiltonian} tensors in the full-MO space
but defining occupied- and virtual-MO slices that include only the active MOs.
This also allows the MO-response components of the derivative codes to
incorporate frozen-core-MO/active-occupied-MO responses in a seamless manner.

\subsubsection{Derivatives and DerivStore}

The calculation of most molecular properties requires derivatives of wave
function amplitudes, MO coefficients, and Hamiltonian one-electron,
two-electron and overlap integrals.\cite{Yamaguchi94}  The \code{Derivatives}
object provides the latter as a wrapper for Psi4's \code{MintsHelper} object's
derivative integral routines.  It provides skeleton first and second
derivatives of the integrals (the derivatives of the AO-basis quantities), as
well as overlap half-derivatives (for AATs, VG-APTs, and DBOCs), and dipole
derivatives (for LG-APTs).  Certain first-derivative quantities are
transformed into the MO basis by the \code{MintsHelper}
functions,\cite{Turney12:PSI4} but for gradients the first-derivative
two-electron integrals and for Hessians the second-derivative two-electron
integrals are computed in the AO basis and contracted directly against
AO-basis quantities (\textit{e.g.}, back-transformed two-electron densities)
to minimize the number of costly basis transformations.  (Orbital-response
contributions are provided by the \code{CPHF} object described below.)

In order to avoid re-computation of the derivatives of four-index tensors,
such as \code{ERI} or two-electron densities, the \code{DerivStore} object
provides persistent HDF5 storage keyed by \code{(quantity, perturbation,
context)}, one per \code{Wavefunction}.  Each time such a derivative tensor is
needed, the call is passed through a \code{get\_or\_compute()} function that
checks for the quantity in the \code{DerivStore} object.  If it exists, the
function reads the tensor from disk and returns it to the caller; if it does
not, then it (a) calls a \code{builder} function provided by the caller to
generate it, (b) stores it on the \code{DerivStore} in case of future use, and
(c) returns the tensor to the caller.  We choose a ``lazy'' approach in which
\code{Derivatives} and \code{DerivStore} (as well as the \code{CPHF} object
below) are created only on first access so that workflows that do not require
the derivative infrastructure avoid \code{h5py} and the creation of an
unnecessary scratch file.  On the first write request, the \code{DerivStore}
checks the type of disk provided and warns the user if a RAM-backed filesystem
is in use, since the cache would consume memory instead of disk.

\subsubsection{CPHF}

Second-order properties such as nuclear Hessians, polarizabilities, and
LG-APTs require the response of the MOs.  The \code{CPHF} object solves the
equations defining the first-order orbital response to an external
perturbation, $x$,
\begin{displaymath}
G_{iajb} U^x_{jb} = B^x_{ia},
\end{displaymath}
\textit{i.e.}, the coupled-perturbed Hartree-Fock (CPHF) equations.  The MO
Hessian, $\mathbf{G}$, \textit{e.g.}, for RHF orbitals,
\begin{displaymath}
G_{iajb} = \left(\epsilon_a - \epsilon_i\right) \delta_{ij} \delta_{ab} + L_{ajib} \pm L_{abij}
\end{displaymath}
is independent of the perturbation, apart from the sign between the two
$L_{pqrs}$ components, which depends on whether the perturbation is real ($+$) or
imaginary ($-$).  The structure of the vector, $B^x_{ia}$, depends on the
nature of the perturbation, with current allowed choices of \code{'field'},
\code{'nuclear'}, \code{'magnetic'}, or \code{'vecpot'} (for a spatially
uniform vector potential that depends on the linear-momentum operator).  This
generalized interface provides an intentional separation between the
orbital-response solver and the property tensors.

The \code{CPHF} object avoids re-computation of perturbation-dependent
four-index tensors such as $\partial_x \langle pq | rs \rangle$ by caching
them in the \code{DerivStore} upon creation.  Certain two-index derivative
quantities are stored in an in-core cache (a Python dictionary), such as the
skeleton Fock and overlap derivatives, the full Fock-matrix derivatives, and
$U^x_{ia}$ CPHF coefficients so that multi-property workflows may re-use them as
needed.  Furthermore, the object provides the full set of $U^x$ coefficients
(that is, including occupied-occupied and virtual-virtual mixings) for
orbital-response contributions to higher-order derivatives with selection
between canonical- and non-canonical perturbed MOs.

\subsection{Derived Wavefunction Classes}

The base class currently feeds into four derived classes, \code{HFwfn},
\code{MPwfn}, \code{CIwfn}, and \code{CCwfn}, as depicted in
Fig.~\ref{fig:wfn_base_class}.  Each subclass inherits all components of
\code{Wavefunction}, including the \code{Hamiltonian}, \code{DeviceManager},
\code{Derivatives}, and \code{CPHF} components, but also adds its own
method-specific quantities and functions.  For example, the \code{MPwfn}
object includes the energy denominators, first-order $\hat{T}_2$ amplitudes
(and first-order $\hat{T}_1$ for semicanonical ROHF references) needed for MP2
calculations.  The \code{CIwfn} and \code{CCwfn} objects include their
respective singles- and doubles-amplitude/coefficient tensors as well as corresponding
\code{solve\_ci} and \code{solve\_cc} functions to compute them, and the
\code{CCwfn} object provides additional model selection across
CCD/CC2/CCSD/CCSD(T)/CC3.

All three correlated wave function objects are instantiated directly by the
user without first creating a base \code{Wavefunction}.  The constructor for
each derived wave function class uses the built-in Python function
\code{super()} that returns a proxy object, which invokes the parent
\code{Wavefunction} class's constructor.  This ensures that all inherited
attributes are correctly initialized and allows the individual wave functions
to handle their own user-input keywords and pass the rest (\textit{e.g.}, the
choice of device or arithmetic precision) to the parent
\code{Wavefunction}'s constructor.

The correlated wave function classes are designed only to provide ground-state
energies and amplitudes.  Other components (\textit{e.g.}, excitation energies
and densities) are contained within downstream classes that are typically
specific to the type of wave function (for example, the \code{ccdensity}
class).  For analytic derivative properties, each wave function class feeds
its own driver (\code{MPderiv}, \code{CIderiv}, and \code{CCderiv}), each of
which is derived from the more general \code{CorrelatedDerivs} parent
class (and which is utilized similarly to the \code{Wavefunction} class above
through the \code{super()} proxy).  \code{CorrelatedDerivs} provides
a method-agnostic interface to the computation of a range of energy-derivative
properties, such as nuclear gradients and Hessians, dipole moments and
polarizabilities, and others.  Unlike the correlated wave functions, however,
the \code{HFwfn} object carries no wave function of its own, since Psi4
provides all the relevant SCF information for \code{Wavefunction.ref}.
Instead \code{HFwfn} is the home for the Hartree-Fock analytic-derivative
property functions, which provide the reference-determinant's contributions to
their correlated counterparts.  These features are described in detail in
section \ref{sec:derivs}.

 \section{Correlated Wave Function Classes}
\label{sec:wfns}

PyCC currently includes three correlated wave function classes: \code{MPwfn},
\code{CIwfn}, and \code{CCwfn}.  The first two were added most recently,
primarily to support our development of simulations of vibrational circular
dichroism (VCD) spectra.  The \code{CCwfn} class was developed in the early
stages of the PyCC package and feeds into multiple additional functions beyond
those available for MP2 and CI.

\subsection{The CCwfn Class}

The fundamental purpose of the \code{CCwfn} class is to compute the
ground-state $\hat{T}$ amplitudes and energies required for the CCD, CC2,
CCSD, CCSD(T), and CC3 models with both spatial- and spin-orbital formulations
available for the latter three models.  Thus, the primary user-facing function
is \code{solve\_cc()}, which iterates the amplitude residuals to
user-specified convergence criteria, with acceleration provided by a general
\code{helper\_diis} object that is used across all iterative wave function classes.  The
CC intermediates used in the residuals are the standard set based on the
similarity transformed Hamiltonian, $\bar{H} = e^{-\hat{T}} \hat{H}
e^{\hat{T}}$, with model-specific definitions as needed.  In addition, the CC3
method adds its own $\hat{T}_1$-transformed intermediates and $\hat{T}_3$
residual.  A local-correlation helper class (\code{Local}, section
\ref{sec:local}) may be triggered using a \code{local} keyword that selects
from PNO, PAO, PNO++, or CPNO++ methods.

Iterative and non-iterative triples contributions are computed using the
now-conventional algorithms for batching $t_{ijk}^{abc}$ amplitudes. The most
effective approach requires looping over $(i,j,k)$ combinations and computing
$v^3$ tensors for all $(a,b,c)$ combinations, but PyCC provides the inverse of
this algorithm (looping over $(a,b,c)$ combinations and computing the $o^3
(i,j,k)$ tensors) for special cases (\textit{e.g.}, CC3 response functions).
For RHF references, the (T) energy correction to CCSD may be computed using the efficient
algorithm developed by Rendell, Lee, and Komornicki,\cite{Rendell91} or by a
simpler, but slower approach that separately computes the second-order $\hat{T}_3$
contributions to third-order $\hat{T}_1$ and $\hat{T}_2$.

\begin{figure*}[!hbtp]
\centering
\includegraphics[width=\textwidth]{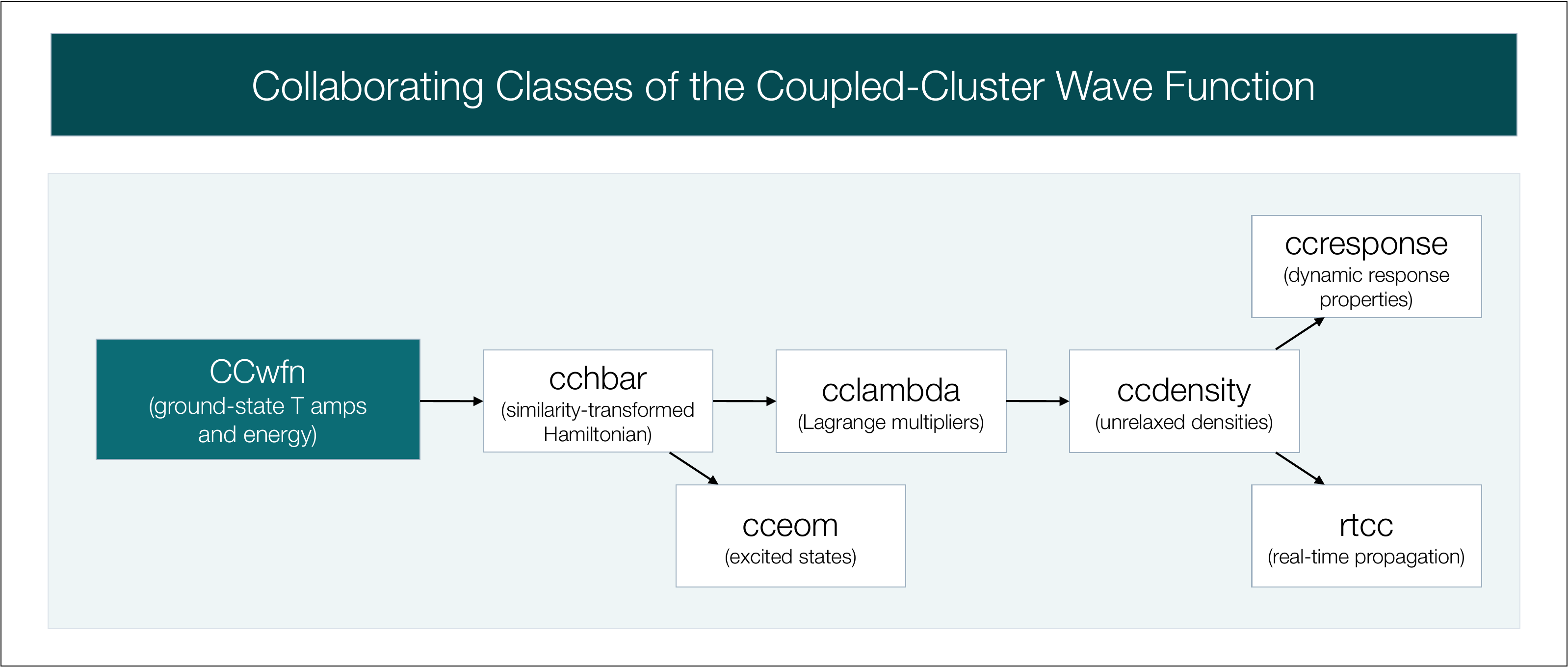}
\caption{Collaborating classes of \code{CCwfn}, including the
similarity-transformed Hamiltonian (\code{cchbar}), the ground-state Lagrange
multipliers (\code{cclambda}), electronically excited states (\code{cceom}),
the ground-state one- and two-electron densities (\code{ccdensity}), dynamic
response functions (\code{ccresponse}), and real-time coupled cluster
propagation (\code{rtcc}).}  \label{fig:ccwfn} \end{figure*}

Beyond ground-state energies, the \code{CCwfn} object feeds into a chain of
collaborating classes, as shown in Fig.~\ref{fig:ccwfn}.  After \code{CCwfn}
has computed the ground-state $\hat{T}$ amplitudes and energy, these are fed
to the \code{cchbar} class, which computes the various components of the
similarity-transformed Hamiltonian, $\bar{H}$.  These may then be fed either
into a \code{cclambda} object for computing the Lagrange multipliers or into a
\code{cceom} object for computing excited states (see section \ref{sec:eom}).
The \code{cclambda} object is then fed to a \code{ccdensity} object, which
prepares the one- and two-electron unrelaxed densities.  These may then be
used to compute one- and two-electron properties, dynamic response functions
(\code{ccresponse}, section \ref{sec:response}), and/or real-time
propagation (\code{rtcc}, section \ref{sec:rtcc}).

\subsection{The MPwfn and CIwfn Classes}

As noted above, the purpose of the \code{MPwfn} and \code{CIwfn} classes is
somewhat simpler than \code{CCwfn}.  Both objects are intended primarily to
compute their respective ground-state wave function amplitudes and energies,
but they also feed into the \code{MPderiv} and \code{CIderiv} objects
described later in section \ref{sec:derivs} for derivative-based properties.
The \code{MPwfn} is currently limited to the first-order MP wave function,
which is computed non-iteratively, and the corresponding MP2
energy.  The \code{CIwfn}, on the other hand, must compute its amplitudes
iteratively (with convergence acceleration from the \code{helper\_diis}), and
we have chosen to use a projected form of the CI equations rather than the
conventional eigenvalue form,\cite{Scuseria88} because it is consistent with
the implementation in \code{CCwfn}.  Note that the \code{CIwfn} is fully
re-normalized after the amplitudes are obtained (and the MP1 wave function is
similarly re-normalized when used in subsequent VCD calculations).
 \section{Analytic Derivatives and Molecular Properties}
\label{sec:derivs}

\begin{figure*}[!hbtp]
\centering
\includegraphics[width=\textwidth]{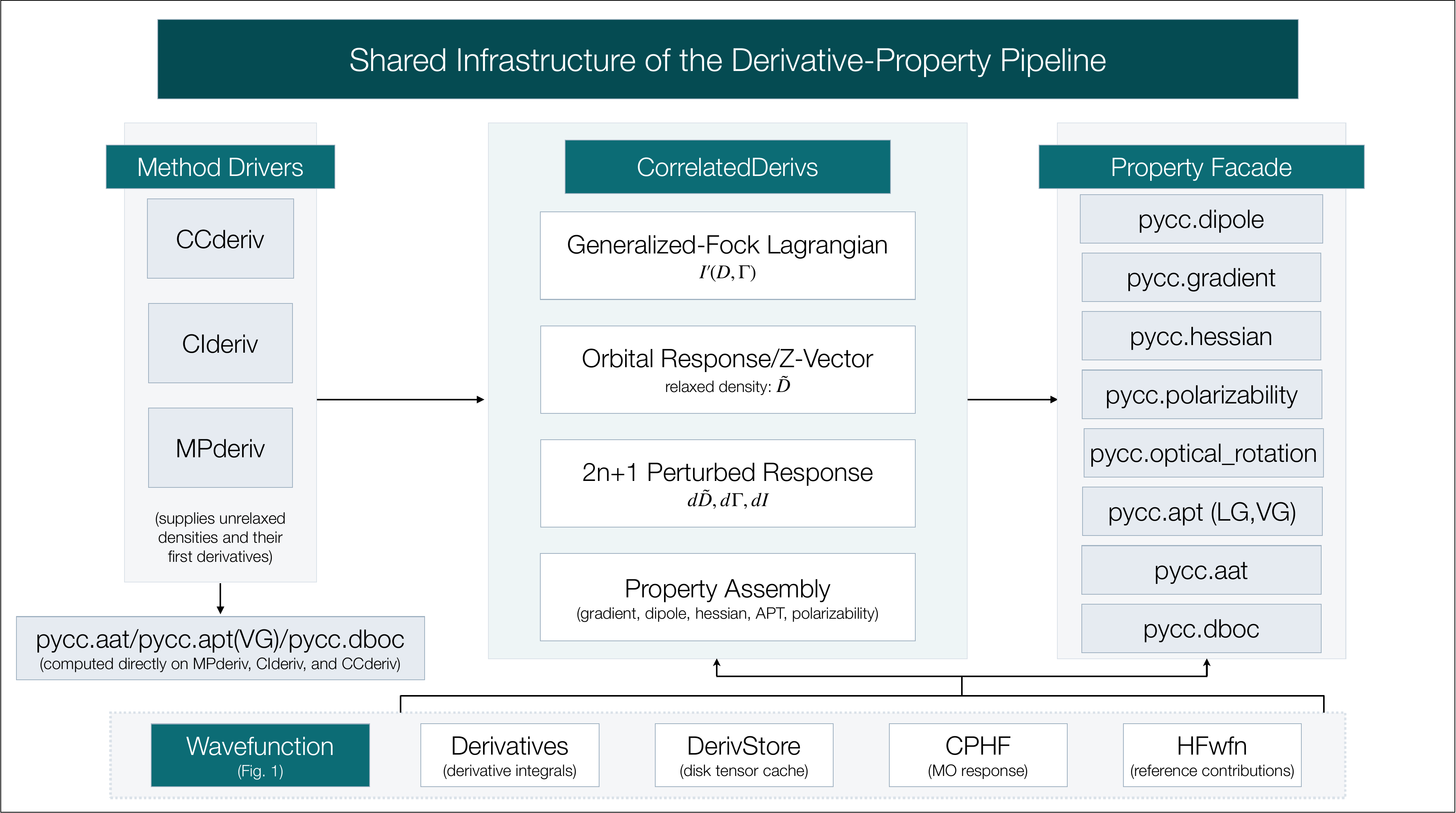}
\caption{Derivative-Property Pipeline: Method-specific classes (\code{CCderiv}, \code{MPderiv}, and \code{CIderiv})
provide unperturbed and perturbed densities to the \code{CorrelatedDerivs} class, which computes the required first-
and/or second-derivatives of the energy for a given property.  For properties such as AATs, VG-APTs, and DBOCs, which
are not formulated as energy derivatives, method-specific functions are provided}
\label{fig:derivs} \end{figure*}

As noted in Table \ref{tab:derivatives}, PyCC provides infrastructure for
computing a range of properties as first and second analytic derivatives, both
using energy derivatives and overlaps of wave function derivatives.  The
former includes molecular gradients and Hessians, dipole moments, static
polarizabilities, and LG-APTs, while the latter include DBOCs, AATs, and
VG-APTs.  The software components that handle such calculations are outlined
schematically in Fig.~\ref{fig:derivs}, which traces the computational flow
from method-specific upstream classes (\code{CCderiv}, \code{MPderiv}, and
\code{CIderiv}), to a method-agnostic \code{CorrelatedDerivs} class, to the
user-facing property interfaces.

\subsection{The CorrelatedDerivs Class}

All energy first- and second-derivative properties are computed by a
\code{CorrelatedDerivs} object, which is based on a density formulation that
is independent of the specific choice of model.\cite{Yamaguchi94, Adamowicz84,
Scheiner87, Gauss91:UHF, Lee91, Scuseria91:grads, Stanton97:Hessians}  This approach takes advantage of the Wigner
$2n+1$ rule for energy derivatives, \textit{i.e.}, the correlated wave
function and MO responses are required starting with second derivatives. 

First derivatives of the correlation energy with respect to an external
parameter $x$ may be written as (in spin-orbital notation),
\begin{equation}
\frac{\partial E}{\partial x} = \sum_{pq} \tilde{D}_{pq} f^{(x)}_{pq} + \sum_{pqrs}\Gamma_{pqrs}\aint{pq}{rs}^{(x)}
+ \sum_{pq} I_{pq} S^{(x)}_{pq},
\label{eq:grad}
\end{equation}
where $\tilde{D}_{pq}$ is the MO-relaxed one-particle density, $\Gamma_{pqrs}$
is the unrelaxed two-particle density, $I_{pq}$ is the generalized Fock
matrix, and $f^{(x)}_{pq}$, $\aint{pq}{rs}^{(x)}$, and $S^{(x)}_{pq}$ are the
``skeleton derivatives'', that is, derivatives of the AO-basis Fock matrix,
two-electron integrals, and overlap matrix transformed into the MO basis.
Thus, given the unrelaxed one- and two-particle densities for a given method
(MP2, CI, CC), \code{CorrelatedDerivs} uses the \code{Derivatives},
\code{DerivStore}, and \code{CPHF} objects that are available on the
\code{Wavefunction} base class to compute the MO $z$-vector,\cite{Handy84}
$I_{pq}$, and $\tilde{D}_{pq}$, and subsequently assemble all the components
in Eq.~\eqref{eq:grad}.  The specific expressions for $\tilde{D}_{pq}$ depend
on the choice of canonical- vs.\ non-canonical perturbed MOs.  The latter uses
overlap derivatives in place of dependent-MO-pair CPHF contributions as
described by Handy and co-workers,\cite{Handy85:MP2Hessians} and the former is
used only for CCSD(T) properties in PyCC following the approach described by
Lee and Rendell.\cite{Lee91} For the latter, numerical instabilities in the
derivatives can arise in cases of exact or near degeneracies of the MOs, as
noted by Lee and Rendell, which are typically handled by choosing cutoffs for
either orbital-rotation gradients or orbital-energy differences below which
the corresponding contributions are set to zero, as discussed in Ref.\ \onlinecite{Lee91}.

Note that the two-electron contributions in Eq.~\eqref{eq:grad} are computed
in the AO basis rather than the MO basis using the conventional approach of
back-transforming $\Gamma_{pqrs}$ rather than transforming the AO-basis
$\aint{pq}{rs}^{(x)}$ for every perturbation.  This equation applies to
general external perturbations, such as nuclear coordinates and static
electric fields (for relaxed dipole moments), though for the latter only the
first term is needed and $f^{(x)}_{pq}$ collapses to the electric-dipole
integrals.  In addition, spatial and spin-orbital approaches are handled
separately with the latter called as needed within a top-level function for
each component (\textit{e.g.}, a \code{\_lagrangian} function handles the RHF
approach by default, but routes to an \code{\_so\_lagrangian} function for
UHF/ROHF cases).

For energy second derivatives, PyCC uses the $2n+1$ rule for wave function
derivatives but not the $2n+2$ rule for derivatives of the Lagrange
multipliers.  This results in an ``asymmetric'' formulation of the second
derivative expression, \textit{viz.},
\begin{align}
\frac{\partial^2 E}{\partial x\,\partial y}
&= \sum_{pq}\tilde D_{pq}f^{(xy)}_{pq}
+ \sum_{pqrs}\Gamma_{pqrs}\aint{pq}{rs}^{(xy)} \nonumber \\
& \quad + \sum_{pq}I_{pq}S^{(xy)}_{pq} + (\text{orbital response}) \nonumber \\
& \quad
+ \sum_{pq}\frac{\partial\tilde D_{pq}}{\partial y}f^{(x)}_{pq}
+ \sum_{pqrs}\frac{\partial\Gamma_{pqrs}}{\partial y}\aint{pq}{rs}^{(x)} \nonumber \\
& \quad + \sum_{pq}\frac{\partial I_{pq}}{\partial y}S^{(x)}_{pq},
\label{eq:hessian}
\end{align}
where the skeleton second-derivatives are defined similarly to
Eq.~\eqref{eq:grad}.  The ``orbital response'' contributions depend on the
choice of canonical- vs.\ non-canonical-perturbed MOs, and, as for the first
derivatives, the two-electron skeleton second-derivative contributions are
computed in the AO basis to avoid repeated costly AO-to-MO transformations of
the derivative integrals.  Like the first derivatives, spatial and
spin-orbital approaches are handled separately, but route through identical
top-level functions.  In addition, we note that static polarizabilities, which
are second derivatives with respect to external electric fields, require only
the contributions from $\tilde{D}_{pq}$ and $\partial_x \tilde{D}_{pq}$.
Mixed second derivatives such as the LG-APT may be formulated either as
nuclear derivatives of the relaxed electric dipole or as field derivatives of
the nuclear gradient.  The latter is the default in PyCC for stand-alone
calculations of the LG-APT because they require computation of perturbed wave
function amplitudes only for the three Cartesian electric-field components.
However, the most common use of the APTs is for simulating IR and VCD spectra.
Such calculations typically compute the vibrational Hessian first, and thus
the $3N$ nuclear-coordinate derivatives of the wave function and its
corresponding densities are already available for the LG-APT on the
\code{DerivStore}, allowing the LG-APT computation at little additional cost.
Thus, in the PyCC functions \code{ir()} and \code{vcd()} which automate such
simulations, the default is overridden in favor of differentiation of the
relaxed dipole moment with respect to the $3N$ nuclear coordinates.

As pointed out by Gauss and Stanton,\cite{Gauss96} while the use of canonical
perturbed-MOs is essential for computing CCSD(T) second derivatives
efficiently, they also can lead to numerical instabilities in the resulting
properties in the case of (near-)degeneracies among the MOs. This can occur
specifically for cases in which the direct product of the irreducible
representations of the degenerate MOs is the same as that of the perturbation
(for $D_{2h}$ and its subgroups).  This problem can be seen in table S1 of the
SM, specifically for UHF-CCSD(T)/cc-pVDZ second derivatives of $^3B_1$
\ce{CH2}, where comparisons of PyCC results against CFOUR exhibit larger
errors than when using non-canonical-perturbed MOs.  Although we have not yet
implemented the approach for avoiding these instabilities described by Gauss
and Stanton in their implementation of CCSD(T) second derivatives for magnetic
shielding constants, this will be included in a future release of PyCC.  In
the present version of the code, however, an option for choosing
non-canonical-perturbed MOs for CCSD(T) first and second derivatives is
provided that yields much better validation against CFOUR properties
(cf.\ Table S1).  While this approach is less efficient than for
canonical-perturbed MOs, our algorithm makes full use of batched generation of
triple-excitation quantities, both avoiding storage of full six-index tensors
and retaining overall ${\cal O}(N^7)$ scaling.

\subsection{Method-Specific Derivative Drivers}

Similar to the relationship between the \code{Wavefunction} base class and the
method-specific wave function classes, the upstream method-specific objects
(\code{MPderiv}, \code{CIderiv}, and \code{CCderiv}) are themselves derived from
\code{CorrelatedDerivs}.  Each takes a converged wave function as input to its
constructor and calls \code{super().\_\_init\_\_()} to obtain all the
attributes of the \code{CorrelatedDerivs} class.  Then each
\code{\{MP,CI,CC\}deriv} builds its own unrelaxed densities for gradients and
both unperturbed and perturbed unrelaxed densities for second derivatives.
The \code{CorrelatedDerivs} infrastructure handles all MO response
contributions and assembly of the requested properties regardless of method.

The \code{MPderiv} class is the simplest because its unrelaxed densities
require only its closed-form (non-iterative) $\hat{T}_2$ amplitudes, and thus
its perturbed doubles, which then yield its unrelaxed perturbed densities, are
obtained non-iteratively from the perturbed integrals and Fock matrix.  
The \code{CIderiv} class is more complicated because both singles and doubles
are needed in CISD, and thus its two-particle density involves all
occupied-MO/virtual-MO combinations.  Furthermore, both unperturbed and
perturbed wave function amplitudes must be computed iteratively.

The \code{CCderiv} class uses much of the pipeline shown in
Fig.~\ref{fig:ccwfn} in that the computation of the CC-level unrelaxed
densities requires $\bar{H}$ and $\hat{\Lambda}$ amplitudes.  Similarly, the
CC perturbed unrelaxed densities require derivatives of both $\hat{T}$ and
$\hat{\Lambda}$ amplitudes, which themselves require derivatives of $\bar{H}$.  For
CCSD(T) the triples contributions to the $\hat{\Lambda}$ equations and the
unrelaxed density are computed at the same time as the (T) energy, and the
derivatives of the (T) density components are computed separately. In
addition, the \code{CCderiv} object can also produce
\textit{frequency-dependent} linear response properties using the same
\code{CorrelatedDerivs} infrastructure.  This will be discussed later in
section \ref{sec:response}.

Properties that are formulated not as energy derivatives but as overlaps of
wave function derivatives, such as the AAT, VG-APT, and DBOC, are directly
computed on their derivative objects.  At present, these quantities are
available for MP2, CI, and CC wave functions in PyCC (see Table
\ref{tab:derivatives}), so the \code{MPderiv}, \code{CIderiv}, and
\code{CCderiv} classes provide them directly.

\subsection{Property Facade}

In order to simplify the user interface, standard properties may be requested
through a general set of functions, including \code{dipole}, \code{gradient},
\code{polarizability}, \code{optical\_rotation}, \code{hessian}, \code{apt},
\code{aat}, and \code{dboc}.  Each function requires a \code{\{MP/CI/CC\}deriv} object and
not just a \code{\{MP/CI/CC\}wfn} object for two reasons.  The first is to
avoid cases in which, for example, the user re-computes the wave function
amplitudes using different convergence criteria after the creation of
\code{\{MP/CI/CC\}deriv}.  The latter creates a snapshot of the converged wave
function (including $\hat{\Lambda}$ in the case of \code{CCderiv}), thus
protecting the property calculations from changes in the state of the original
wave function parameters that would cause a mismatch with densities and other
components computed from the previous set of converged parameters.

The second reason for requiring a \code{\{MP/CI/CC\}deriv} object for the
property interface is to ensure consistent use of the \code{DerivStore}
disk-caching system.  Each of the perturbed tensors in the \code{DerivStore}
is identified by the \code{CorrelatedDerivs} object that owns it.  This
prevents two or more \code{CorrelatedDerivs} objects (\textit{e.g.} an
\code{MPderiv} and a \code{CCderiv}) existing simultaneously from accidentally
accessing the same \code{DerivStore} tensors.

Each property function returns a \code{PropertyComponents} object that
includes the nuclear (if any), Hartree-Fock reference, and correlation
contributions to the property, and the interface to the property returns the
electronic and/or total property as sums of the appropriate components.  These
quantities are all computed separately, \textit{e.g.}, the correlation
component is not obtained as a difference between the reference and electronic
components.
 \section{Coupled Cluster Response Methods}
\label{sec:response}

PyCC provides two independent implementations of dynamic (frequency-dependent)
coupled cluster linear response functions, the first-order response of a given
property such as the electric- or magnetic-dipole moment to a time-dependent
external field.  For example, the linear response of the electric dipole to an
external electric field is proportional to the dipole polarizability, while
the linear response of the same dipole to an external magnetic field is
proportional to the $G'$ (or optical activity) tensor.  As was noted in section
\ref{sec:derivs}, one of these implementations makes use of the second
derivative infrastructure of \code{CorrelatedDerivs}, and therefore uses the
same density-based second-derivative expressions needed for static
polarizabilities, vibrational Hessians, and LG-APTs.  An essential difference
between the computation of such static properties and dynamic response
functions is that the latter are typically computed without explicit inclusion
of orbital response, \textit{i.e.}, the MOs are not permitted to relax in
response to the external fields in order to avoid spurious poles at finite
fields.  This is handled by the \code{CCderiv} object by simply computing only
the unrelaxed one- and two-electron (unperturbed and perturbed) densities in
place of their relaxed counterparts.  In addition, the field enters the
amplitude equations as the bare MO (electric or magnetic) dipole integrals
with no two-electron components and the frequency ($\omega$) of the external
field appears as a scalar shift in the inhomogeneous terms in the residuals
and in the orbital-energy denominators.

The second formulation lives in the \code{ccresponse} class, which explicitly
implements the CC diagrammatic expressions for the various components of the
response function in terms of $\bar{H}$ matrix elements, unperturbed
$\hat{\Lambda}$ amplitudes, and the field-perturbed $\hat{T}$ amplitudes.
Thus, a key difference between the formulations in \code{ccresponse} and
\code{CCderiv/CorrelatedDerivs} is that the former follows both the $2n+1$ and
$2n+2$ rules for the perturbed wave function amplitudes, requiring only the
derivatives of the $\hat{T}$ amplitudes with respect to the field (the
``symmetric'' derivative formulation), while the latter follows only the
$2n+1$ rule and thus requires field derivatives of both $\hat{T}$ and
$\hat{\Lambda}$ amplitudes (the ``asymmetric'' formulation).  This does not
reduce the number of perturbed wave function equations that must be solved
because in the symmetric approach in \code{ccresponse}, one must still compute
the derivatives of $\hat{T}$ with respect to both $+\omega$ and $-\omega$,
rather than derivatives of both $\hat{T}$ and $\hat{\Lambda}$ with respect to
only $+\omega$ (and both sets of equations have the same polynomial scaling).
However, the symmetric approach is still somewhat less expensive than the
asymmetric route because the expressions for the perturbed $\hat{\Lambda}$
amplitudes involve significantly more tensor contractions, resulting in a
larger prefactor.  Both implementations yield identical results, which provides
a validation check between the two distinct formulations.

Both implementations provide CCSD-level dynamic linear response functions for
RHF, UHF, and ROHF references (\textit{i.e.} spatial- and spin-orbital-MO
formulations), whereas CC3 linear response is currently implemented only for
spin-orbitals in the \code{ccresponse} code.  The \code{ccresponse}
implementation allows for six different one-electron perturbation operators,
including the electric-dipole moment, the magnetic-dipole moment (and its
complex conjugate), the linear-momentum operator (and its complex conjugate),
and the traceless quadrupole operator; \code{CCderiv} permits only the
electric- and magnetic-dipole operators.  The corresponding perturbed wave
functions are computed only when needed, and thus a polarizability computation
requires only an electric-dipole-moment perturbation (three Cartesian
directions with $\pm\omega$ or both $\hat{T}/\hat{\Lambda}$ = six total
perturbed amplitude sets), whereas an optical rotation computation requires
both electric- and magnetic-dipole perturbations plus the complex conjugate of
the latter (12 total perturbed amplitude sets).  While only linear response
functions are presently available in PyCC, a pull-request for quadratic
response, which requires the derivatives of the $\hat{\Lambda}$ amplitudes, is
under review and will be added in a future release.
 \section{Equation-of-Motion Coupled Cluster}
\label{sec:eom}

PyCC provides excitation energies and wave functions through the
equation-of-motion (EOM-CC) formalism,\cite{Stanton93:EOM} in which the
eigenvalues and eigenvectors of the similarity-transformed Hamiltonian,
$\bar{H}$, are computed using a generalized block-Davidson-Liu algorithm.  As
shown in Fig.~\ref{fig:ccwfn}, the \code{cceom} class requires only
\code{cchbar} and the underlying \code{CCwfn} as input, and, at present,
only the EOM-CCSD level of theory is supported (for RHF, UHF, or ROHF
references).  The single user-facing function in \code{cceom} is
\code{solve\_eom}, and users may choose to compute either right- or left-hand
excitation vectors, with support for computing multiple states simultaneously.
In each iteration of the algorithm, all $N$ roots are expanded as a block, with
per-root convergence locking.  A cutoff (\code{root\_floor}) in a.u.\ is used to filter
the Ritz values such that a root is accepted only if it is real
($\left|\mathrm{Im}\right| < 10^{-6}$ a.u.) and its real part is higher than the floor
(default value of $10^{-3}$).  These features provide improved stability,
particularly for near-degenerate states and spurious near-zero roots.  In
addition, three types of guess vectors are permitted: unit vectors (selected
based on single-excitation differences in diagonal components of the
one-electron matrix elements of $\bar{H}$), eigenvectors of the configuration
interaction singles (CIS) Hamiltonian, and the eigenvectors of the
singles-singles block of $\bar{H}$.  At present, no properties other than
EOM-CCSD excitation energies are available, though a future release will
include analytic derivatives and CC3.
 \section{Real-Time Coupled Cluster}
\label{sec:rtcc}

PyCC includes a submodule containing the \code{rtcc} class, which allows time-dependent propagation of the CC right- and
left-hand wave functions in the presence of an external electric field through the equations
\begin{equation}
\dot{t}_\mu = -i \bra{\mu} \bar{H} \ket{0}
\label{eq:derivT}
\end{equation}
and
\begin{equation}
\dot{\lambda}_\mu = i \bra{0} \left(1 + \hat{\Lambda}\right) \left[\bar{H}, \tau_\mu\right] \ket{0}
\label{eq:derivL}
\end{equation}
where $\tau_\mu$ is a second-quantized operator for the $\mu$-th excitation. The propagation is carried out by casting
these coupled ordinary differential equations into the general form
\begin{equation}
\dot{y}(t) = f(y,t),
\end{equation}
where $y(t)$ is a concatenation of the $\hat{T}$ and $\hat{\Lambda}$
amplitudes at a given time step (the left-hand sides of Eqs.~\eqref{eq:derivT}
and \eqref{eq:derivL}), and $f(y,t)$ is the pair of residuals on the right-hand
sides.  In each step of the propagation, which is handled by the
\code{rtcc.propagate()} function, the $\hat{T}$ and $\hat{\Lambda}$ amplitude
tensors are flattened into a single one-dimensional vector ($y(t)$) and passed
into the (creatively named) function \code{rtcc.f()}.  This function
extracts the amplitudes, adds to the Fock matrix the value at time $t$ of the
external field (whose form is chosen by the user among five pre-defined laser
envelopes, including a Dirac-delta kick), computes the $\hat{T}$ and
$\hat{\Lambda}$ residuals from the corresponding functions in \code{CCwfn} and
\code{cclambda} (multiplied by $-i$ and $+i$, respectively), computes the
phase from the total CC energy ($\dot{\phi} = -i E_\textrm{CC}$), and
re-flattens the amplitudes (and the phase) into a new vector $y$, which is
returned to the caller.  The \code{rtcc.propagate()} function then uses the
user-selected integrator (either chosen from PyCC's built-in set of nine
integrators or provided by the user externally) to take a step ($h$), and
computes the desired properties (typically the electric- and/or magnetic-dipole
moments) from the density at that point in time.  The propagation continues as
long as the user has requested, and the resulting dipole trajectory is
typically used to compute an absorption or circular dichroism spectrum via
Fourier transformation of the appropriate real or imaginary component.
Utilities are provided for post-propagation analysis including a fast Fourier
transform (FFT), FFT power-spectrum thresholding, damping via an exponential
envelope, and Pad\'{e}-accelerated spectra.  An advantage of real-time methods
over their response theory counterparts is that the simulations may be
adjusted to more closely mimic experiments, \textit{e.g.} by adjusting the
definition of the laser field or even using multiple
fields.\cite{Krishnan2026}

Note that the propagation equations Eqs.~\eqref{eq:derivT} and
\eqref{eq:derivL} are complex (through the time-dependent Schr\"odinger
equation), and thus the resulting time-dependent amplitudes are complex.  A
key aspect of the PyCC \code{rtcc} implementation is that the complex
arithmetic required is handled by the \code{ContractionBackend} object, which
upcasts the operands to complex so that real Hamiltonian integrals contract
against complex time-dependent amplitudes with no changes to the CC residual
or density equations (see section \ref{subsec:wfn}).  In addition, the
user-selected single- (complex64) or double-precision (complex128), as well as
GPU/CPU options, are similarly handled automatically by
\code{ContractionBackend}.  These features were reported and tested in our
work on RT-CC3.\cite{Wang2025}

Another important feature of the structure of \code{rtcc} is that the same
residual equations from \code{CCwfn} and \code{cclambda} serve both the
ground-state amplitude solvers and the RTCC propagation, so that any method
added to the ground-state codes automatically enjoys access to the
time-dependent framework.  The only method-specific branching exceptions in
\code{rtcc} are for the RT-CCD quasienergy (phase) and for the density and
select special intermediates for RT-CC3.  The \code{rtcc} object also computes
and returns the autocorrelation function (the overlap of the CC wave functions
at an initial time $t_0$ and current time step), which can be useful for an
MO-decomposition of the absorption peaks in the spectra.\cite{Krishnan2026} In
addition, checkpointing is handled automatically by \code{rtcc.propagate()} so
that long propagations may be re-started after a crash (or batch queue time
limits).  Finally, local correlation such as LPNO-CC may be explored in the
RT-CC propagation using a simulation algorithm in which the $\hat{T}$ and
$\hat{\Lambda}$ residuals are filtered to remove non-local
components.\cite{Peyton2023}  This will be explained further in section
\ref{sec:local}.
 \section{Local Coupled Cluster}
\label{sec:local}

PyCC has proven its worth for rapid-prototyping and testing of new methods,
especially in the area of local-correlation/reduced-scaling CC
methods.\cite{Peyton2023, Madriaga2025:disc, Madriaga2025}  The package has two
distinct local-CC execution modes: a ``simulation'' approach that uses the
canonical-MO code to filter non-local contributions iteration-by-iteration and
a genuine pairwise (localized occupied-MO) algorithm that more directly mimics
a fully optimized code.  The PyCC local-CC code provides four potential
virtual-orbital spaces: (1) PAOs\cite{Saebo93, Hampel96, Schutz01}, the AO
basis with the localized occupied MO contributions projected out --- one
overall virtual space with subspaces selected on a per-occupied-MO-pair basis;
(2) PNOs,\cite{Neese09} pair natural virtual orbitals constructed separately
for each localized occupied-MO pair; and (3) PNO++/CPNO++,\cite{Crawford19,
Dcunha2021, Madriaga2025} PNOs replaced or augmented, respectively, by
contributions from the perturbation-dependent pair densities.  These spaces are
selected and deployed through a \code{Local} class which handles the
construction of the localized virtual spaces, as well as the relevant
transformation matrices.  The occupied MOs are localized by the base
\code{Wavefunction} class using either the Pipek-Mezey\cite{Pipek89} or
Foster-Boys\cite{Boys60} approaches, and the local-CC codes are limited to
spatial MOs only at present.

The local-simulation code within the \code{Local} class uses an approach
originally described by Hampel and Werner\cite{Hampel96} and works through the
canonical-MO CC codes in PyCC.  In each iteration (of, \textit{e.g.}, the
ground-state CC equations) the residual's virtual-MO space is first
transformed to the local space (PAO, PNO, PNO++/CPNO++) for each pair and then
transformed to a semicanonical space for which the virtual-MO Fock matrix for
each pair is diagonal.  The orbital-energy denominator in that representation
is applied to the residual to filter out non-local components and the
resulting increment is transformed back to the canonical-MO space and applied
to the previous iteration's amplitudes.  While this approach produces the same
results as a full pairwise algorithm, it offers no improvement in memory and
CPU scaling.  Thus, its value is primarily in prototyping and validation, and
it is straightforwardly extended to canonical-MO codes with minor code
additions (\textit{e.g.}, the PyCC RTCC\cite{Peyton2023} and response
codes\cite{Madriaga2025}).

The pairwise algorithm on the other hand, provides a more complete comparison
to an optimized local-correlation approach by restructuring all of the relevant
equations and their in-code implementation to use only fully localized
components using a \code{lccwfn} class.  This would require a complete
reformulation of all residuals, intermediates, $\bar{H}$ components, energy
expressions, and response functions (some of which have been implemented in
pull requests that are currently under review).  While the present Python
implementation cannot be considered optimal by any means, it matches the
results of the simulation route and should serve as a validation source for
truly production-level implementations.
 \section{Validation, Documentation, and Testing}
\label{sec:docs}

PyCC validates its results against four distinct types of reference data: (1)
external production codes such as Psi4\cite{Smith20} or
CFOUR;\cite{Matthews2020:CFOUR} (2) high-order numerical derivatives of PyCC's
own energies or first-derivative quantities; (3) internal cross-checks between
two routes to the same quantity (\textit{e.g.}, spatial- vs.\ spin-orbital
formulations); and (4) exact identities that the given property must satisify
(\textit{e.g.}, sum rules or comparison between CISD and CCSD for two-electron
systems).  The last category is particularly important when there is no
existing external validation source available, such as the newly developed
CC-level AATs.\cite{Crawford2026:ccvcd}  Table S1 of the Supplementary
Material reports the measured deviation of the PyCC CC-level analytic
derivative properties for each validation reference (as opposed to the test
tolerance used in the test suite).  The largest deviations ($6.2\times
10^{-10} \text{\ to\ } 3.3\times10^{-8}$) occur for three open-shell CCSD(T)
second-derivative properties (Hessian, static polarizability, and LG-APT) and
arise due to numerical instabilities arising from the use of
canonical-perturbed MOs. For all other test cases the highest deviation ($3.5
\times 10^{-10}$ a.u.) occurs only for cases for which the external source
provides limited precision, such as CFOUR Hessians.

The PyCC documentation consists of two components: (1) the in-code docstrings
that specify both the class and function interfaces/attributes as well as the
details of the relevant mathematical expressions; and (2) working theory and
design documents that provide the foundational physics and code-implementation
plans that have guided PyCC development over the years, but especially since
June 2026. The in-code docstrings present the mathematics in two formats, a
plain-text version that is readable in a terminal and a \code{.. math::} form
for rendered output, with the latter especially important for the \code{CCwfn},
\code{Derivatives}, \code{cchbar}, \code{CPHF}, \code{CorrelatedDerivs}, and
\code{ccresponse} classes.  A Sphinx build renders the docstrings in
browsable API documentation accessible via, \textit{e.g.}, Read the Docs.

The theory and design documents are included as tracked files in the PyCC
GitHub repository.  The former include detailed derivations formatted as LaTeX
documents (\textit{e.g.}, \code{cc\_gradients\_orbital\_response.tex}, which
lays out the details of the first- and second-derivative infrastructure
implemented in \code{CorrelatedDerivs}).  The design documents provide both
design plans and an historical record of the day-to-day development of the
code.

The testing infrastructure of PyCC currently spans ca.\ 100 test files
comprising about 450 test functions spanning nearly 90\% of the source code.
These tests are executed in relevant batches when specific features or
debugging efforts are carried out, but the suite is run when more major
developments are implemented, pull requests are opened/modified, or new code
is merged into the \code{main} branch.  In accord with current software
engineering best practices, the latter two are executed automatically by
continuous integration scripts on the fully open-source GitHub repository, and
the current suite requires approximately 45 minutes to execute (excluding
particularly long tests that are used for validation and occasional regression
checks).  The continuous integration testing utilizes two operating system
types (MacOS and Linux) and two dependency configurations (with and without
PyTorch).  Given that standard continuous integration runners do not provide
GPU access, the PyTorch routes in PyCC are tested against their CPU fallbacks.

 \section{Conclusions and Prospectus}
\label{sec:conclusions}

PyCC is a Python-based code whose purpose is fourfold: (1) to deliver a clear
reference implementation of a range of quantum chemical wave-function-based
methods; (2) to provide a validation suite for the development of large-scale
production-level implementations in packages such as Psi4; (3) rapid
prototyping of new ideas, such as local-CC based on PNO++ and CPNO++
perturbation-aware correlation spaces or RT-CC3; and (4) to contribute to the
education of neophytes in the basic algorithms of quantum chemical models.  The
code is fully open source, meaning that its entire development history (as well
as current efforts) are publicly exposed at its GitHub repository.\cite{PyCC:code}

PyCC is \textbf{not} intended to provide a production-level, optimized
computational infrastructure for large molecules.  While its algorithms do not
fully duplicate advanced infrastructure, they do include important efficiency
measures such as disk-based storage of quantities that are expensive to
compute and store in core memory, as well as AO-based approaches in derivative
calculations that minimize the costly AO-to-MO transformation of
perturbation-dependent quantities.  Thus, PyCC reaches beyond ``toy'' problems
into realistic test cases that can serve to stress emerging models.  For
example, our recent efforts include simulation of the VCD spectrum of
3-chloro-1-butyne at the CCSD/cc-pVTZ level of theory (224 contracted Gaussian
functions, frozen core).  The calculation included the fully analytic
vibrational Hessian and LG-APT, as well as the analytic AAT and VG-APT, which
are only available at the MP2, CI, and CC levels of theory in
PyCC.\cite{Shumberger2024, Shumberger2025, Shumberger2026, Krishnan2026:density, Crawford2026:ccvcd}
The full computation required 21 hours on 12 cores of an AMD 7763 CPU plus 272
GiB of memory and roughly 1.8 TiB of local storage for PyCC's
\code{DerivStore}. The peak memory usage is dominated by the AO-basis
second-derivative integral blocks, which are produced by \code{Psi4} by pairs
of atoms ($9 n_\text{AO}^4 \cdot 8$ bytes).  The \code{DerivStore} consists
almost entirely of $n_\text{MO}^4$ tensors approximately evenly divided among
skeleton MO-basis derivative two-electron integrals (with no orbital
response), total MO-basis derivative two-electron integrals, and derivative
(relaxed) two-electron densities, one for each nuclear perturbation.
Magnetic- and linear-momentum perturbations add ca.\ 6\%, and perturbed
cluster amplitudes add less than 1\%.  The total footprint of the
\code{DerivStore} is about $(9N+6) n_\text{MO}^4 \cdot 8$ bytes.  Our
experience is that calculations on the order of 300 basis functions for
molecules of comparable size to 3-chloro-1-butyne are routinely feasible
provided sufficient memory and disk resources.

PyCC is fully open source and available at GitHub with the BSD 3-Clause
license.\cite{PyCC:code} New contributions to the code are welcomed and follow
current best-practices for collaborative software development such as pull
requests from branches or forks, careful review by a select number of human
core developers, continuous integration testing, etc.

Our near- and long-term goals for PyCC include, but are not limited to: (1)
production-level versions of most PyCC capabilities within the Psi4
infrastructure as enabled by the emerging Einsums C++ contraction backend; (2)
quadratic response functions (with the pull request under review); (3)
coupled-cluster AATs and VG-APTs for more advanced VCD spectral simulations;
(4) EOM properties and transition strengths; (5) pairwise local $\hat{\Lambda}$
and $\bar{H}$ (with pull requests under review); (6) open-shell local CC; and
(7) broader GPU coverage.

In closing, while the initial design and foundational code development of PyCC
began in 2021 based entirely on hand-written code, documentation, and testing,
more recent development (since ca.\ June 2026) of the package has made
use of the AI agent Claude Code (Anthropic, running the Claude Opus
4.8, 5, and 5.5 models), especially the extensive implementation of the
analytic derivative infrastructure.
 Our philosophy and best practices for the use of such tools emphasize
careful human guidance, oversight, and review of \textbf{all} AI products.
Most of the work is guided by externally prepared theory documents written by
the developers or earlier versions of hand-written external code that the AI
uses as a template for deployment within PyCC.  The validation efforts
described earlier include external references (both established codes and
numerical derivatives), internal consistency checks, and comparison to exact
identities, and thus add to careful human guidance to guard against AI errors.
Our view is that the AI agent is helpful in carrying out tasks that are
already well-understood but slow when implemented by a human, but the agent is
unhelpful and precarious when used for tasks without a complete and clear
advance specification. While we consider the use of an AI agent for
documentation and test-case construction to be perhaps the safest and most
straightforward use of AI in code development in the computational sciences
because such contributions do not directly touch source code, extensive
external validation tests remain essential.
 
\section{Acknowledgements} \label{ack}

This work was supported by the U.S.\ National Science Foundation via grants
CHE-2154753 and OAC-2410880.  The authors are grateful to Advanced Research
Computing at Virginia Tech for providing computational resources that have
contributed to the results reported within the paper.  The authors wish to
acknowledge invaluable assistance by Thomas Pedersen and H\a kon Kristiansen
in validating the PyCC RT-CC codes in its early stages of development.

\end{document}


\title{Supplementary Material\\
PyCC: A Reference Quantum Chemistry Python Package for Education, Validation,
and Rapid Prototyping}

\author{T. Daniel Crawford}
	\affiliation{Department of Chemistry, Virginia Tech, Blacksburg, VA 24061, U.S.A.}
	\email{crawdad@vt.edu}
\author{Benjamin G. Peyton}
	\affiliation{Department of Chemistry, Virginia Tech, Blacksburg, VA 24061, U.S.A.}
\author{Zhe Wang}
	\affiliation{Department of Chemistry, Virginia Tech, Blacksburg, VA 24061, U.S.A.}
\author{Aparna Krishnan}
	\affiliation{Department of Chemistry, Virginia Tech, Blacksburg, VA 24061, U.S.A.}
\author{Brendan M. Shumberger}
	\affiliation{Department of Chemistry, Virginia Tech, Blacksburg, VA 24061, U.S.A.}
\author{Jose Madriaga}
	\affiliation{Department of Chemistry, Virginia Tech, Blacksburg, VA 24061, U.S.A.}
\author{Joshua Atta-Kumi}
	\affiliation{Department of Chemistry, Virginia Tech, Blacksburg, VA 24061, U.S.A.}
\author{Aaron De Clercq}
	\affiliation{Department of Chemistry, Virginia Tech, Blacksburg, VA 24061, U.S.A.}
\author{Ritvik Prabhu}
	\affiliation{Department of Chemistry, Virginia Tech, Blacksburg, VA 24061, U.S.A.}
\date{\today}

\maketitle

\footnotesize
\setlength{\LTcapwidth}{\textwidth}
\begin{longtable}{lllllllr}
\caption{Measured agreement between PyCC and its validation references. Each entry is $\max|\mathrm{PyCC}-\mathrm{reference}|$ in atomic units. All-electron and frozen-core cases are listed separately. Agreement is the measured result, not the assertion tolerance used in the test suite. ``NCP'' in the Route column marks rows computed with non-canonical perturbed molecular orbitals; the remaining CCSD(T) rows use the canonical perturbed-MO gauge, which is PyCC's default for that method. Lettered notes follow the table. \newline\textit{Alt text: table of the measured agreement between PyCC and each validation reference for 77 cases, spanning Gradient, Hessian, APT (length gauge), Static polarizability and AAT for CCSD and CCSD(T). Agreement is at the $10^{-14}$ to $10^{-10}$ level throughout, apart from 3 open-shell CCSD(T) second-derivative rows reaching $10^{-8}$, where the canonical perturbed-MO gauge is ill-conditioned.}}\label{tab:validation}\\
\hline\hline
Property & Method & Molecule & Basis & Route & Case & Reference & Agreement \\
\hline
\endfirsthead
\multicolumn{8}{l}{\textsc{Table}~\thetable\ (continued)}\\
\hline\hline
Property & Method & Molecule & Basis & Route & Case & Reference & Agreement \\
\hline
\endhead
\hline
\multicolumn{8}{r}{\footnotesize\itshape continued on the next page}\\
\endfoot
\hline\hline
\endlastfoot
Gradient & CCSD & \ce{CH2} & cc-pVDZ & spin-orbital & all-electron & CFOUR\textsuperscript{a,e} & $4.2\times10^{-11}$ \\*
 & CCSD(T) & \ce{CH2} & 6-31G & spin-orbital, NCP & all-electron & CFOUR\textsuperscript{a,e} & $2.8\times10^{-11}$ \\*
 & CCSD(T) & \ce{CH2} & cc-pVDZ & spin-orbital & all-electron & CFOUR\textsuperscript{a,e} & $4.6\times10^{-11}$ \\*
 & CCSD(T) & \ce{H2O} & 6-31G & spatial & all-electron & Finite difference\textsuperscript{b,f} & $1.8\times10^{-12}$ \\*
 & CCSD(T) & \ce{H2O} & 6-31G & spatial & frozen core & Finite difference\textsuperscript{b,f} & $1.9\times10^{-12}$ \\
\hline
Hessian & CCSD & \ce{H2O} & 6-31G & spatial & all-electron & CFOUR\textsuperscript{a,e} & $2.6\times10^{-10}$ \\*
 & CCSD & \ce{H2O} & 6-31G & spatial & frozen core & CFOUR\textsuperscript{a,e} & $2.7\times10^{-10}$ \\*
 & CCSD & \ce{H2O} & 6-31G & spin-orbital & all-electron & CFOUR\textsuperscript{a,e} & $2.6\times10^{-10}$ \\*
 & CCSD & \ce{H2O} & 6-31G & spin-orbital & frozen core & CFOUR\textsuperscript{a,e} & $2.7\times10^{-10}$ \\*
 & CCSD & \ce{H2O} & cc-pVDZ & spatial & all-electron & CFOUR\textsuperscript{a,e} & $2.2\times10^{-10}$ \\*
 & CCSD & \ce{H2O} & cc-pVDZ & spatial & frozen core & CFOUR\textsuperscript{a,e} & $1.8\times10^{-10}$ \\*
 & CCSD & \ce{H2O} & cc-pVDZ & spin-orbital & all-electron & CFOUR\textsuperscript{a,e} & $2.2\times10^{-10}$ \\*
 & CCSD & \ce{H2O} & cc-pVDZ & spin-orbital & frozen core & CFOUR\textsuperscript{a,e} & $1.8\times10^{-10}$ \\*
 & CCSD & \ce{H2O+} & STO-3G & spin-orbital & all-electron & CFOUR\textsuperscript{a,e} & $1.2\times10^{-10}$ \\*
 & CCSD & \ce{H2O+} & STO-3G & spin-orbital & frozen core & CFOUR\textsuperscript{a,e} & $1.3\times10^{-10}$ \\*
 & CCSD(T) & \ce{CH2} & 6-31G & spin-orbital & all-electron & CFOUR\textsuperscript{a,e} & $9.1\times10^{-11}$ \\*
 & CCSD(T) & \ce{CH2} & 6-31G & spin-orbital, NCP & all-electron & CFOUR\textsuperscript{a,e} & $9.1\times10^{-11}$ \\*
 & CCSD(T) & \ce{CH2} & cc-pVDZ & spin-orbital & all-electron & CFOUR\textsuperscript{a,g} & $\sim6.2\times10^{-10}$ \\*
 & CCSD(T) & \ce{CH2} & cc-pVDZ & spin-orbital, NCP & all-electron & CFOUR\textsuperscript{a,e} & $1.2\times10^{-10}$ \\*
 & CCSD(T) & \ce{H2O} & 6-31G & spatial & all-electron & CFOUR\textsuperscript{a,e} & $1.4\times10^{-10}$ \\*
 & CCSD(T) & \ce{H2O} & 6-31G & spatial & frozen core & CFOUR\textsuperscript{a,e} & $1.7\times10^{-10}$ \\*
 & CCSD(T) & \ce{H2O} & 6-31G & spin-orbital & all-electron & CFOUR\textsuperscript{a,e} & $1.4\times10^{-10}$ \\*
 & CCSD(T) & \ce{H2O} & cc-pVDZ & spatial & all-electron & CFOUR\textsuperscript{a,e} & $1.2\times10^{-10}$ \\*
 & CCSD(T) & \ce{H2O} & cc-pVDZ & spatial & frozen core & CFOUR\textsuperscript{a,e} & $3.5\times10^{-10}$ \\
\hline
APT (length gauge) & CCSD & \ce{H2O} & 6-31G & spatial & all-electron & CFOUR\textsuperscript{a,e} & $2.2\times10^{-10}$ \\*
 & CCSD & \ce{H2O} & 6-31G & spatial & frozen core & CFOUR\textsuperscript{a,e} & $1.8\times10^{-10}$ \\*
 & CCSD & \ce{H2O} & 6-31G & spin-orbital & all-electron & CFOUR\textsuperscript{a,e} & $2.2\times10^{-10}$ \\*
 & CCSD & \ce{H2O} & 6-31G & spin-orbital & frozen core & CFOUR\textsuperscript{a,e} & $1.8\times10^{-10}$ \\*
 & CCSD & \ce{H2O} & cc-pVDZ & spatial & all-electron & CFOUR\textsuperscript{a,e} & $6.4\times10^{-11}$ \\*
 & CCSD & \ce{H2O} & cc-pVDZ & spatial & frozen core & CFOUR\textsuperscript{a,e} & $1.3\times10^{-10}$ \\*
 & CCSD & \ce{H2O} & cc-pVDZ & spin-orbital & all-electron & CFOUR\textsuperscript{a,e} & $6.4\times10^{-11}$ \\*
 & CCSD & \ce{H2O} & cc-pVDZ & spin-orbital & frozen core & CFOUR\textsuperscript{a,e} & $1.3\times10^{-10}$ \\*
 & CCSD & \ce{HOF} & 6-31G & spatial & all-electron & CFOUR\textsuperscript{a,e} & $1.7\times10^{-10}$ \\*
 & CCSD & \ce{HOF} & 6-31G & spatial & frozen core & CFOUR\textsuperscript{a,e} & $1.0\times10^{-10}$ \\*
 & CCSD(T) & \ce{CH2} & 6-31G & spin-orbital & all-electron & CFOUR\textsuperscript{a,e} & $1.1\times10^{-10}$ \\*
 & CCSD(T) & \ce{CH2} & 6-31G & spin-orbital, NCP & all-electron & CFOUR\textsuperscript{a,e} & $1.1\times10^{-10}$ \\*
 & CCSD(T) & \ce{CH2} & cc-pVDZ & spin-orbital & all-electron & CFOUR\textsuperscript{a,g} & $\sim3.6\times10^{-9}$ \\*
 & CCSD(T) & \ce{CH2} & cc-pVDZ & spin-orbital, NCP & all-electron & CFOUR\textsuperscript{a,e} & $6.4\times10^{-11}$ \\*
 & CCSD(T) & \ce{H2O} & 6-31G & spatial & all-electron & CFOUR\textsuperscript{a,e} & $1.2\times10^{-10}$ \\*
 & CCSD(T) & \ce{H2O} & 6-31G & spatial & frozen core & CFOUR\textsuperscript{a,e} & $1.6\times10^{-10}$ \\*
 & CCSD(T) & \ce{H2O} & 6-31G & spin-orbital & all-electron & CFOUR\textsuperscript{a,e} & $1.2\times10^{-10}$ \\*
 & CCSD(T) & \ce{H2O} & 6-31G & spin-orbital & frozen core & CFOUR\textsuperscript{a,e} & $1.6\times10^{-10}$ \\*
 & CCSD(T) & \ce{H2O} & cc-pVDZ & spatial & all-electron & CFOUR\textsuperscript{a,e} & $1.2\times10^{-10}$ \\*
 & CCSD(T) & \ce{H2O} & cc-pVDZ & spatial & frozen core & CFOUR\textsuperscript{a,e} & $1.7\times10^{-10}$ \\*
 & CCSD(T) & \ce{HOF} & 6-31G & spatial & all-electron & CFOUR\textsuperscript{a,e} & $7.0\times10^{-11}$ \\*
 & CCSD(T) & \ce{HOF} & 6-31G & spatial & frozen core & CFOUR\textsuperscript{a,e} & $1.3\times10^{-10}$ \\
\hline
Static polarizability & CCSD & \ce{CH2} & cc-pVDZ & spin-orbital & all-electron & CFOUR\textsuperscript{a,e} & $3.7\times10^{-11}$ \\*
 & CCSD & \ce{H2O} & cc-pVDZ & spatial & all-electron & CFOUR\textsuperscript{a,e} & $4.7\times10^{-11}$ \\*
 & CCSD & \ce{H2O} & cc-pVDZ & spatial & frozen core & CFOUR\textsuperscript{a,e} & $3.0\times10^{-11}$ \\*
 & CCSD & \ce{H2O} & cc-pVDZ & spin-orbital & all-electron & CFOUR\textsuperscript{a,e} & $4.6\times10^{-11}$ \\*
 & CCSD & \ce{H2O} & cc-pVDZ & spin-orbital & frozen core & CFOUR\textsuperscript{a,e} & $3.0\times10^{-11}$ \\*
 & CCSD(T) & \ce{CH2} & 6-31G & spin-orbital & all-electron & CFOUR\textsuperscript{a,e} & $4.3\times10^{-11}$ \\*
 & CCSD(T) & \ce{CH2} & 6-31G & spin-orbital, NCP & all-electron & CFOUR\textsuperscript{a,e} & $4.3\times10^{-11}$ \\*
 & CCSD(T) & \ce{CH2} & cc-pVDZ & spin-orbital & all-electron & CFOUR\textsuperscript{a,g} & $\sim3.3\times10^{-8}$ \\*
 & CCSD(T) & \ce{CH2} & cc-pVDZ & spin-orbital, NCP & all-electron & CFOUR\textsuperscript{a,e} & $5.0\times10^{-11}$ \\*
 & CCSD(T) & \ce{HOF} & 6-31G & spatial & all-electron & CFOUR\textsuperscript{a,e} & $6.1\times10^{-11}$ \\*
 & CCSD(T) & \ce{HOF} & 6-31G & spatial & frozen core & CFOUR\textsuperscript{a,e} & $5.4\times10^{-11}$ \\*
 & CCSD(T) & \ce{H2O} & 6-31G & spatial & all-electron & CFOUR\textsuperscript{a,e} & $4.8\times10^{-11}$ \\*
 & CCSD(T) & \ce{H2O} & 6-31G & spatial & frozen core & CFOUR\textsuperscript{a,e} & $4.6\times10^{-11}$ \\*
 & CCSD(T) & \ce{H2O} & 6-31G & spin-orbital & all-electron & CFOUR\textsuperscript{a,e} & $4.8\times10^{-11}$ \\*
 & CCSD(T) & \ce{H2O} & 6-31G & spin-orbital & frozen core & CFOUR\textsuperscript{a,e} & $4.6\times10^{-11}$ \\*
 & CCSD(T) & \ce{H2O} & cc-pVDZ & spatial & all-electron & CFOUR\textsuperscript{a,e} & $4.0\times10^{-11}$ \\*
 & CCSD(T) & \ce{H2O} & cc-pVDZ & spatial & frozen core & CFOUR\textsuperscript{a,e} & $3.5\times10^{-11}$ \\
\hline
AAT & CCSD & \ce{H2O} & cc-pVDZ & spatial & all-electron & magnetic gauge\textsuperscript{c} & $1.7\times10^{-14}$ \\*
 & CCSD & \ce{H2O} & cc-pVDZ & spatial & all-electron & perturbed-MO gauge\textsuperscript{c} & $3.0\times10^{-14}$ \\*
 & CCSD & \ce{H2O} & cc-pVDZ & spatial & frozen core & magnetic gauge\textsuperscript{c} & $1.2\times10^{-14}$ \\*
 & CCSD & \ce{H2O} & cc-pVDZ & spatial & frozen core & perturbed-MO gauge\textsuperscript{c} & $3.2\times10^{-14}$ \\*
 & CCSD & \ce{H2O} & cc-pVDZ & spin-orbital & all-electron & translational sum rule\textsuperscript{c} & $1.5\times10^{-13}$ \\*
 & CCSD & \ce{H2O} & cc-pVDZ & spin-orbital & all-electron & magnetic gauge\textsuperscript{c} & $1.0\times10^{-14}$ \\*
 & CCSD & \ce{H2O} & cc-pVDZ & spin-orbital & frozen core & magnetic gauge\textsuperscript{c} & $1.7\times10^{-14}$ \\*
 & CCSD & \ce{H2O2} & 6-31G & spatial & all-electron & translational sum rule\textsuperscript{c} & $1.9\times10^{-13}$ \\*
 & CCSD & \ce{H2O2} & 6-31G & spatial & all-electron & size intensivity\textsuperscript{c} & $2.5\times10^{-14}$ \\*
 & CCSD & \ce{H2O2} & 6-31G & spatial & frozen core & translational sum rule\textsuperscript{c} & $1.9\times10^{-13}$ \\*
 & CCSD & \ce{H4^2+} & 6-31G & spatial & all-electron & CISD AAT\textsuperscript{c} & $1.2\times10^{-12}$ \\*
 & CCSD & \ce{H4^2+} & cc-pVDZ & spatial & all-electron & CISD AAT\textsuperscript{c} & $1.2\times10^{-12}$ \\*
 & CCSD & \ce{H2O} & cc-pVDZ & --- & all-electron & two routes within PyCC\textsuperscript{d} & $2.9\times10^{-14}$ \\*
 & CCSD & \ce{H2O} & cc-pVDZ & --- & frozen core & two routes within PyCC\textsuperscript{d} & $2.3\times10^{-14}$ \\
\end{longtable}
\normalsize

\noindent\footnotesize\textbf{Notes.}\par\smallskip
\noindent\textsuperscript{a}Analytic CFOUR result: \texttt{VIB=EXACT} for the Hessians and APTs and \texttt{PROP=SECOND\_ORDER} for the polarizabilities, in both cases with \texttt{SCF\_CONV=13}, \texttt{CC\_CONV=12} and \texttt{LINEQ\_CONV=12}.  The 6-31G and STO-3G cases use CFOUR's own GENBAS basis, transcribed into PyCC so the AO bases agree to all printed digits; cc-pVDZ requires no transcription, PyCC's and CFOUR's agreeing to $4\times10^{-13}$ in the SCF energy.\par\smallskip
\noindent\textsuperscript{b}Central finite difference of PyCC's own correlation energy, which shares no density, orbital-response, or perturbed-amplitude machinery with the analytic code.\par\smallskip
\noindent\textsuperscript{c}An identity the quantity must satisfy exactly: two-electron exactness of CCSD against CISD, the translational sum rule, or invariance to a redundant orbital-rotation gauge.\par\smallskip
\noindent\textsuperscript{d}Two independent routes to the same quantity within PyCC (spin-orbital against spin-adapted spatial).\par\smallskip
\noindent\textsuperscript{e}Set by the precision to which the reference provides the data, not by PyCC.\par\smallskip
\noindent\textsuperscript{f}Set by the finite difference, not by PyCC: the residual is the $O(h^4)$ truncation of the stencil.\par\smallskip
\noindent\textsuperscript{g}Set by the conditioning of the canonical perturbed-MO gauge, not by PyCC's formulation: this system has an accidental near-degeneracy between virtual orbitals of different irreducible representations ($\Delta\epsilon = 3.5\times10^{-5}$~$E_\mathrm{h}$), and the canonical gauge divides by that gap.  The value is reproducible within one execution context but shifts by roughly 10\% between them, so it is quoted to one figure.  The non-canonical rows for the same case are unaffected.\par\smallskip
\normalsize